\documentclass{webofc}
\usepackage[varg]{txfonts}   
\usepackage[colorlinks=true,linktocpage=true,linkcolor=blue,citecolor=blue,allcolors=blue]{hyperref}
\begin{document}
\title{QCD at finite temperature and density: Criticality}

\author{\firstname{Volodymyr} \lastname{Vovchenko}\inst{1}\fnsep\thanks{\email{vvovchenko@uh.edu}}
}

\institute{Physics Department, University of Houston, Box 351550, Houston, TX 77204, USA}

\abstract{%
We overview recent theoretical developments in the search for QCD critical point at finite temperature and density, including from lattice QCD, effective QCD theories, and proton number cumulants in heavy-ion collisions.
We summarize the available constraints and predictions for the critical point location and discuss future challenges and opportunities.
}
\maketitle
\section{Introduction}
\label{sec:intro}
Discovering the QCD critical point would establish a major landmark on the phase diagram of strongly interacting matter.
This challenge has motivated many theoretical studies and is one of the primary goals of the beam energy scan program at RHIC~\cite{Bzdak:2019pkr}.
The analytic crossover transition at vanishing net baryon density has long been established from first-principle lattice QCD calculations~\cite{Aoki:2006we}, indicating that the QCD critical point, if it exists, can only be located at finite baryon density.
First-principle constraints on its location mainly come from various extrapolations from $\mu_B = 0$ axis and are thus limited to moderate densities.
Effective QCD theories constrained to lattice data provide guidance into larger baryon densities, albeit with difficult-to-control systematic errors.
Finally, an experimental search for QCD critical point is underway through measurements of cumulants of proton number at different energies, which are expected to reflect the large density fluctuations near the CP.
Several new developments across these different topics were presented at this conference.

\section{Theory estimates}
\label{sec:lqcd}

{\bf Lattice QCD constraints.}
From the theoretical point of view, the existence and location of the QCD critical point should ideally be established from first principles. 
As lattice QCD is the only known first-principle method in the non-perturbative regime, this would necessitate direct lattice simulations at finite $\mu_B$, which are hindered by the sign problem.
Although extremely challenging, it may not be impossible, as discussed by Wong~\cite{Wong}.

Existing constraints instead rely on various extrapolations from $\mu_B = 0$ and imaginary $\mu_B$ simulations that are sign problem free.
One could then look signals of criticality in various observables,
such as the critical lensing of the isentropes or drop in the speed of sound.
Parotto~\cite{Parotto} and Clarke~\cite{Clarke} presented state-of-the-art results on the behavior of these quantities, showing no indications for criticality at moderate values of $\mu_B/T \lesssim 3$.

Further constraints might be obtained by exploiting the expected analytic properties of the QCD partition function near CP.
The CP is a singularity on the real $\mu_B$ axis at $T = T_{\rm CP}$.
At temperatures above $T_{\rm CP}$ this singularity moves into the complex plane, forming a pair of complex conjugate Yang-Lee edge~(YLE) singularities~\cite{Stephanov:2006dn}.
The location of YLE singularities near the CP is governed by universality and given by the appropriate $Z(2)$ scaling,
\begin{align}
\label{eq:LY}
\mu_B^{\rm LY}(T) \simeq \mu_B^{\rm CP} + a (T-T_{\rm CP}) + b (T-T_{\rm CP})^2 + i \, c (T - T_{\rm CP})^{\beta \delta},
\end{align}
where $\beta \delta \simeq 1.56$ is the critical exponent from the 3D-Ising universality class.

\begin{figure}[h]
\centering
\includegraphics[width=0.7\textwidth,clip]{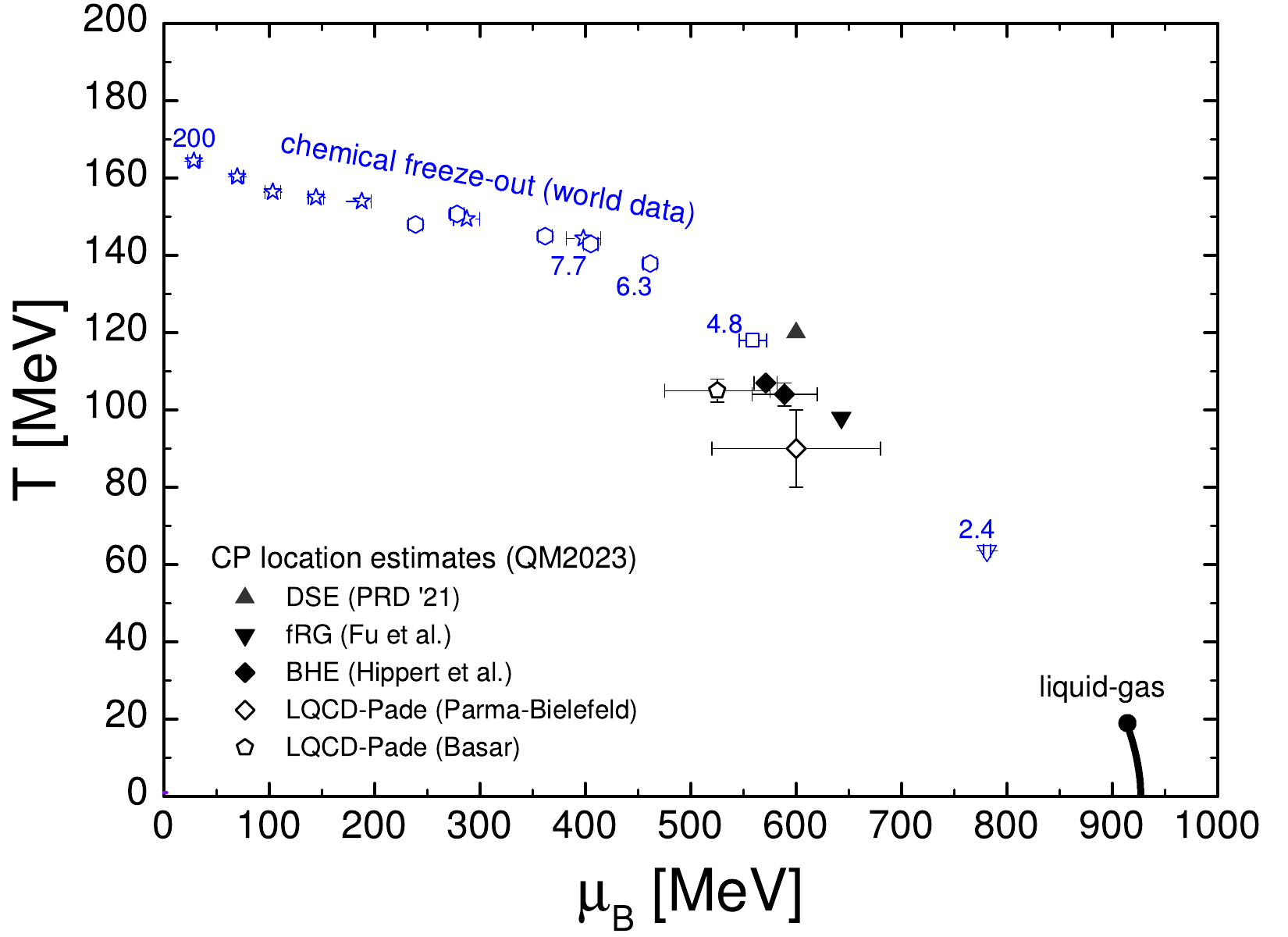}
\caption{The phase diagram of QCD showing recent CP location predictions~\cite{Goswami,Basar,Gunkel:2021oya,Fu:2023lcm,Hippert:2023bel}~(black symbols) and world data on chemical freeze-out points in heavy-ion collisions~\cite{STAR:2017sal,Vovchenko:2015idt,Motornenko:2021nds}~(blue symbols).}
\label{fig:CP}       
\end{figure}

One strategy for searching the CP would be to determine complex plane singularities at several supercritical temperatures, fitting them with Eq.~\eqref{eq:LY} and thus determining the CP location $(T_{\rm CP}, \mu_B^{\rm CP})$.
Complex plane singularities can, in principle, be determined from $\mu_B = 0$ data, for instance, as poles of Pad\'e approximants. 
An accurate determination, however, requires many Taylor expansion coefficients, which so far are only available up to $\chi_8^B$.
Other methods can be used to improve the accuracy, such as multi-point Pad\'e that exploits imaginary $\mu_B$ data or conformal Pad\'e through appropriate mapping of variables.
Goswami~\cite{Goswami} and Basar~\cite{Basar} presented the application of these methods for extracting the complex plane singularities from lattice data and estimating the CP location, reporting $T_{\rm CP} \sim 90-100$~MeV and $\mu_B^{\rm CP} \sim 500-600$~MeV.
These estimates still have large systematic uncertainty as they rely on several assumptions to be valid, namely that Pad\'e approximants determine the complex plane singularity accurately and that it indeed is a LYE singularity in the scaling regime~\eqref{eq:LY} near the QCD CP.
Both assumptions can certainly be questioned at this point, and it can also be noted that the above predictions tend to place the CP below the chemical freeze-out curve from heavy-ion collisions~(Fig.~\ref{fig:CP}), where the CP is not expected to exist.
Nevertheless, this is an encouraging first application of this method in the search for QCD CP, and the control over the systematic uncertainty will likely improve.

{\bf Effective QCD theories.}
Other guidance comes from effective QCD approaches that are not hindered by the sign problem. 
Functional approaches such as Dyson-Schwinger equations or the functional renormalization group have been particularly prominent. These, in principle, allow the mapping of the entire QCD phase diagram, and state-of-the-art calculations~\cite{Gunkel:2021oya,Fu:2019hdw,Fu:2023lcm} indicate the presence of the critical point at $T \sim 100-120$~MeV and $\mu_B \sim 600-650$~MeV. 
These approaches, however, deal with infinite hierarchies of equations and thus require truncations to make calculations tractable. 
It remains to be seen how much influence these approximations have on the CP predictions.

Another interesting development was presented by Hippert~\cite{Hippert:2023bel}, which utilizes an Einstein-Maxwell-Dilaton~(EMD) model based on holographic gauge-gravity correspondence. 
The authors incorporate state-of-the-art lattice QCD data on entropy density and baryon number susceptibility at $\mu_B = 0$ to fix the model input and use Bayesian analysis to find tight constraints on the CP within this approach, placing it at $T \sim 105$~MeV and $\mu_B \sim 580$~MeV.
As with functional methods, the main source of systematic uncertainty lies in how well the EMD model approximates full QCD at finite baryon density.
While the model yields an excellent description of the lattice data at $\mu_B = 0$, that is also true for some approaches, such as the cluster expansion model~\cite{Vovchenko:2017gkg}, that contain no critical point.

{\bf Link to heavy-ion collisions.}
The new predictions for QCD CP location presented at this conference are summarized in Fig.~\ref{fig:CP}.
It is encouraging to see that all the different approaches predict the CP in roughly the same ballpark of $T_{\rm CP} \sim 100-120$~MeV and $\mu_B^{\rm CP}/T_{\rm CP} \sim 5-6$.
To put these predictions in the context of heavy-ion collisions, various chemical freeze-out points are also shown in Fig.~\ref{fig:CP}.
The CP predictions lie close to the freeze-out curve, with the collision energy range of $\sqrt{s_{\rm NN}} \simeq 3-5$~GeV being in closest proximity.
The STAR Fixed Target program and the CBM experiment at FAIR are probing these collision energies.
One caveat here is that the CP predictions in Fig.~\ref{fig:CP} are obtained for vanishing charge and strangeness chemical potentials, $\mu_Q = \mu_S = 0$, whereas in heavy-ion collisions, these are fixed to non-zero values to reproduce the initial charge-to-baryon ratio and strangeness neutrality.
These differences are expected to have a subleading but not necessarily negligible impact.

\section{Critical point search with heavy-ion collisions}

Heavy-ion collisions scan the QCD phase diagram by varying the collision energy, with lower energies corresponding to higher $\mu_B$~(see Fig.~\ref{fig:CP}).
QCD critical point, if it exists in the heavy-ion regime, should manifest itself in observables sensitive to its presence.
Cumulants of baryon number are particularly prominent, as they probe baryochemical potential derivatives of the partition function in equilibrium.
Near a CP, the cumulants exhibit universal singular behavior~\cite{Stephanov:1999zu} and reflect the macroscopically large density fluctuations, well-known in classical systems through the phenomenon of critical opalescence.
The systems created in heavy-ion collisions are too small and short-lived to observe critical opalescence directly, however, the CP signals in (high-order) cumulants can be expected to survive~\cite{Stephanov:2008qz,Kuznietsov:2022pcn}. 

{\bf Theory vs experiment: caveats.}
Analysis and interpretation of experimental measurements of fluctuations can be challenging due to several caveats.

\begin{itemize}
    \item Theoretical predictions for the cumulants are usually made assuming the grand-canonical heat bath. The system in heavy-ion collisions, on the other hand, expands into the vacuum, and the canonical effects due to conservation laws are expected to be sizable~\cite{Bzdak:2012an}. 
    A framework called subensemble acceptance method~(SAM) has been developed recently to make the corresponding corrections~\cite{Vovchenko:2020tsr,Vovchenko:2020gne,Vovchenko:2021yen}, which was discussed in detail by Poberezhnyuk~\cite{Poberezhnyuk}.

    \item Differences between coordinate space, where correlations live, and momentum space accessible in the experiment~\cite{Ling:2015yau,Ohnishi:2016bdf,Asakawa:2019kek}.

    \item Use of measurable proxies~(net-proton) in place of conserved charges (net-baryon)~\cite{Kitazawa:2011wh,Vovchenko:2020kwg}.

    \item Effects of volume and initial state fluctuations, and baryon stopping~\cite{Gorenstein:2011vq,Skokov:2012ds}.

    \item Hadronic phase and non-equilibrium (memory) effects~\cite{Mukherjee:2015swa,Asakawa:2019kek}.
    
\end{itemize}

The above caveats indicate that a dynamical framework for describing fluctuations in heavy-ion collisions is necessary. One can distinguish two different strategies: (i) dynamical modeling of critical fluctuations and (ii) precision calculations of non-critical baseline.

{\bf Modeling of critical fluctuations.}
One way to search for the critical point with heavy-ion collisions is to compare the data with quantitative predictions for proton number cumulants dependent on the assumed location of the CP.
This has been one of the primary goals within the BEST~Collaboration~\cite{An:2021wof}, and involves the development of a generalized hydrodynamic framework to model the non-equilibrium evolution of fluctuations~\cite{Stephanov:2017ghc}, lattice-based QCD equation of state with a tunable critical point~\cite{Parotto:2018pwx,Kapusta:2022pny}, and generalized Cooper-Frye particlization for fluctuations~\cite{Pradeep:2022mkf,Pradeep:2022eil}.
As an alternative to hydrodynamics, especially at lower collision energies, one can consider hadronic transport~\cite{Sorensen:2020ygf} or molecular dynamics~\cite{Kuznietsov:2022pcn} with critical fluctuations.

A new construction of the EoS with a critical point was presented by Kahangirwe~\cite{Kahangirwe}, which utilizes lattice QCD input through the alternative expansion scheme~\cite{Borsanyi:2021sxv} instead of Taylor expansion, extending the applicability of the resulting EoS to the entire BES range. The corresponding equilibrium expectations for proton number cumulants in the presence of QCD CP, as well as other properties of the QCD EoS, were discussed by Karthein~\cite{Karthein}.

New developments on non-equilibirum evolution of non-Gaussian fluctuations in hydrodynamic medium were presented by An~\cite{An}. 
These entail large non-equilibrium memory effects on three- and four-point correlations near the CP due to critical slowing down, which, as shown by Pradeep~\cite{Pradeep},
can not only suppress the signal by more than an order of magnitude but even change the qualitative behavior relative to equilibrium expectations.
In a separate contribution, Savchuk has discussed the effect of diffusion, as well as correlations between QCD conserved charges and energy-momentum~\cite{Savchuk:2023yeh}.
Overall, the modeling of critical fluctuations is not yet on a level applicable for quantitative comparisons with experiment, but steady progress is being made.

{\bf Non-critical baseline.}
Another strategy in searching for the CP lies in analyzing deviations in the data from non-critical expectations.
It has now been established that even the non-critical baseline is a non-trivial function of the collision energy~\cite{Braun-Munzinger:2020jbk}, necessitating precision calculations of essential non-critical contributions to proton number cumulants.
The most prominent contributions are (i) exact global baryon conservation and (ii) repulsive core in the baryon-baryon interaction.
The corresponding calculation was achieved in Ref.~\cite{Vovchenko:2021kxx}, based on (3+1)D relativistic hydrodynamics description of Au-Au collisions at RHIC-BES energies~\cite{Shen:2020jwv} and the appropriately generalized Cooper-Frye particlization routine~\cite{Vovchenko:2022syc}.

Comparison of hydrodynamic calculations with experimental data indicates that (net-)proton number cumulants measured by the STAR Collaboration in 0-5\% central Au-Au collisions are quantitatively consistent with non-critical physics at collision energies $\sqrt{s_{\rm NN}} \gtrsim 20$~GeV, which is also consistent with the measurements at the LHC by the ALICE Collaboration~\cite{ALICE:2019nbs}.
The experimental uncertainties on net proton $\kappa_4/\kappa_2$ are too large to conclude whether a non-monotonic energy dependence relative to a non-critical baseline -- the predicted CP signature~\cite{Stephanov:2008qz} -- is present.
On the other hand, lower-order cumulants such as $\kappa_2$ and $\kappa_3$ are already measured with much better precision and permit additional analysis.

\begin{figure}[h]
\centering
\includegraphics[width=0.49\textwidth,clip]{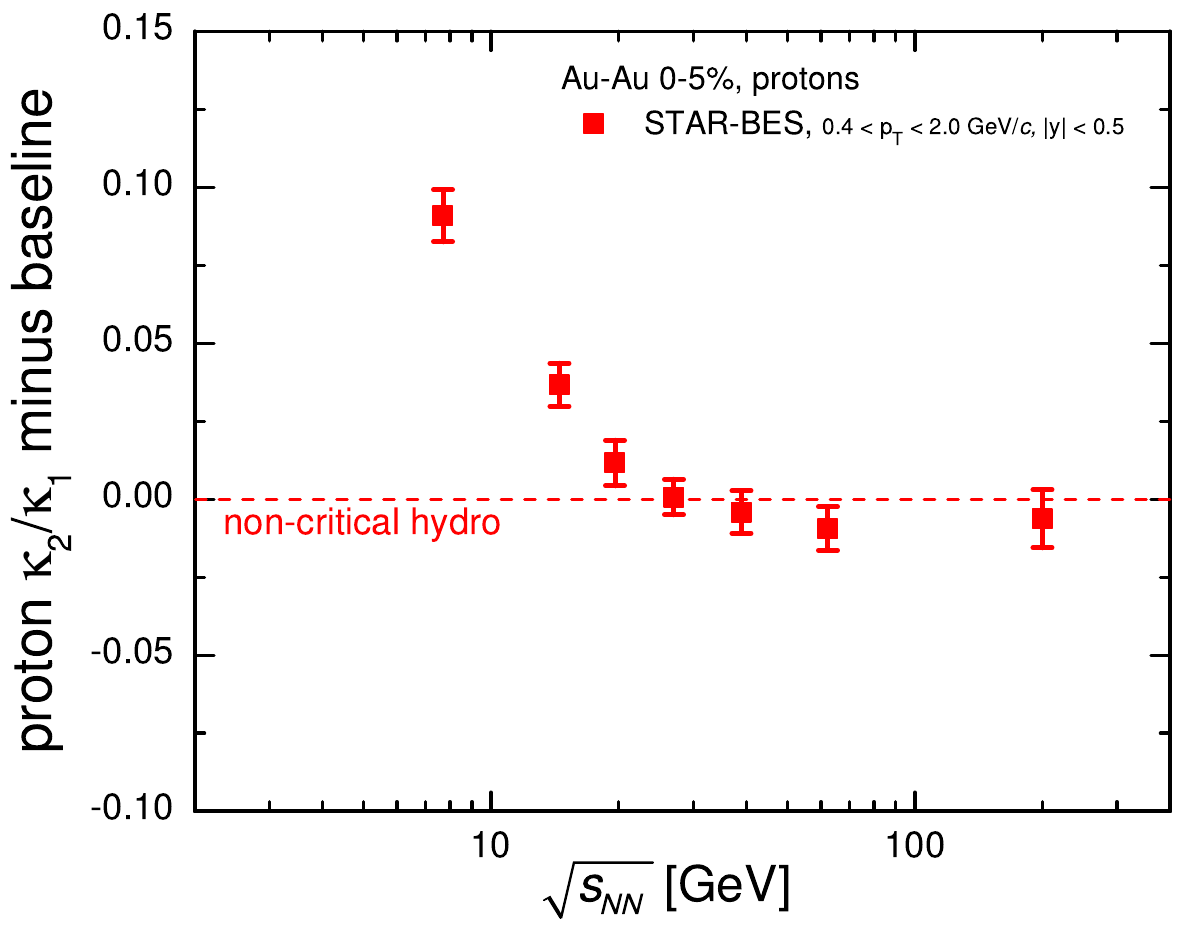}
\includegraphics[width=0.49\textwidth,clip]{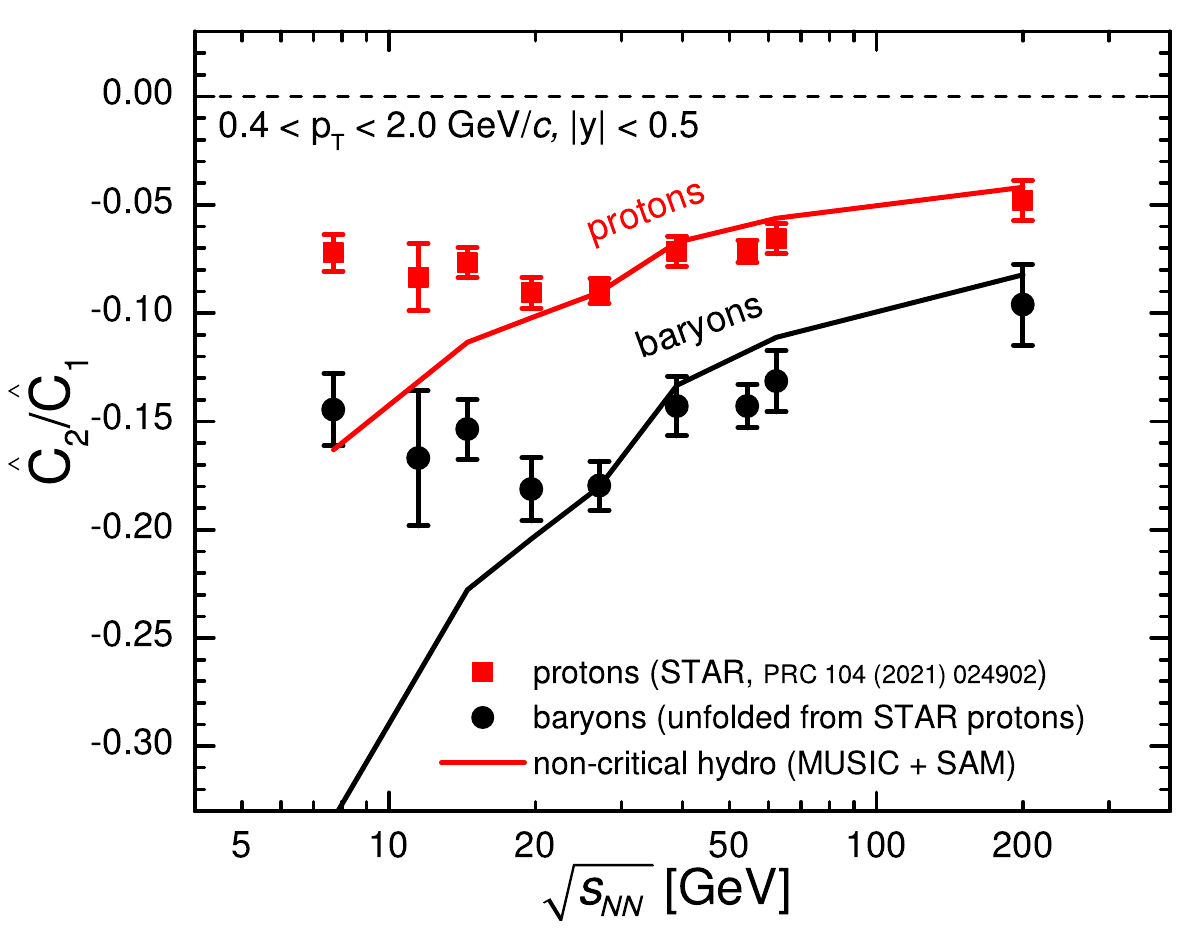}
\caption{Beam energy dependence of (\emph{left panel}) the excess of proton number scaled variance $\kappa_2/\kappa_1$ in 0-5\% central Au-Au collisions~\cite{STAR:2021fge} over non-critical hydrodynamics baseline~\cite{Vovchenko:2021kxx} and (\emph{right panel}) the second scaled factorial cumulant of protons and reconstructed baryons. }
\label{fig:k2k1excess}       
\end{figure}

Figure~\ref{fig:k2k1excess} shows a clear excess of proton number scaled variance $\kappa_2/\kappa_1$ in 0-5\% central Au-Au collisions measured by the STAR Collaboration from BES-I~\cite{STAR:2021iop} over the non-critical hydrodynamics baseline~\cite{Vovchenko:2021kxx} at $\sqrt{s_{\rm NN}} \lesssim 10$~GeV.
This result could be consistent with a CP in baryon-rich matter as predicted by recent effective QCD theories discussed in the previous section.
Available measurements at even lower energies, namely at 3~GeV~(STAR~\cite{STAR:2021fge}) and 2.4~GeV~(HADES~\cite{HADES:2020wpc}), indicate that the excess persists toward lower energies, although an improved control over volume fluctuations and rapidity coverage~\cite{Bzdak} is required for further interpretations.

{\bf Factorial cumulants.}
In addition to ordinary cumulants, $\kappa_n$, factorial cumulants, $\hat{C}_n$, of protons and antiprotons provide complementary information.
These quantities are constructed as linear combinations of cumulants~\cite{Bzdak:2016sxg}, and have the advantage that they remove the trivial~(Poisson) self-correlations and probe genuine multi-particle correlations.
In particular, non-critical effects such as baryon conservation~\cite{Bzdak:2016jxo} or excluded volume~\cite{Vovchenko:2021kxx} are small in high-order factorial cumulants.
On the other hand, high-order factorial cumulants are as singular as ordinary cumulants in vicinity of the CP~\cite{Ling:2015yau}, making them particularly suitable to probe critical behavior. In particular, a non-monotonic collision energy dependence of proton $\hat{C}_4/\hat{C}_1$ could be a more clear CP signature than the commonly discussed $\kappa_4/\kappa_2$.

{\bf Proton vs baryon.}
Due to experimental limitations, fluctuation measurements in heavy-ion collisions are performed on protons rather than on all baryons.
As the critical mode is expected to couple equally to protons and neutrons, the qualitative CP signal should appear in proton cumulants.
However, the magnitude of the correlation signal is affected since, for instance, out of all correlated nucleon-nucleon pairs, only the proton-proton ones show up in the measurement.
Modeling the loss of neutrons as efficiency correction, one can relate proton and baryon scaled factorial cumulants as $\frac{\hat{C}_n^p}{\hat{C}_1^p} \simeq \frac{1}{2^{n-1}} \frac{\hat{C}_n^B}{\hat{C}_1^B}$~\cite{Kitazawa:2011wh}, i.e. a 50\% reduction in the scaled 2nd moment~(Fig.~\ref{fig:k2k1excess}), and a larger loss for high-order moments.
This has to be kept in mind when searching for the CP signal in high-order proton factorial cumulants.
These considerations also indicate that direct quantitative comparisons of net proton cumulant measurements with net baryon cumulant theory calculations~(such as lattice QCD) are unjustified and may lead to misleading conclusions.
At the very least, either the data should be unfolded to estimate net baryon number cumulants, for instance using isospin randomization~\cite{Kitazawa:2011wh}, or the theoretical calculations should be performed for (net) protons directly~\cite{Vovchenko:2021kxx}.

{\bf Other observables.}
Proton number cumulants are not the only observables potentially sensitive to the CP.
Interesting new measurements of azimuthal correlations of protons were presented by Neff~\cite{Neff}, which hint at effective repulsion at all RHIC-BES-II energies.
Measurements of proton intermittency of the NA61/SHINE Collaboration discussed by Porfy~\cite{Porfy} show no structure indicative of a power-law scaling expected from the CP.
The CP and the associated spinodal instability are also expected to induce clustering of particles~\cite{Poberezhnyuk}, which can leave an imprint on light nuclei production~\cite{KJSun}.
Eventually, a consistent understanding of all the observables associated with the criticality will be required.

\section{Summary}

The existence and location of the QCD critical point is still elusive. 
Both the first-principle lattice QCD constraints as well as experimental measurements of proton number cumulants disfavor QCD CP at low baryon densities~($\mu_B/T \lesssim 2-3$) and high collision energies, $\sqrt{s_{\rm NN}} \gtrsim 20$~GeV.
On the other hand, new developments utilizing various effective QCD approaches constrained to lattice data predict a QCD CP in the ballpark of $T_{\rm CP} \sim 100-120$~MeV and $\mu_B^{\rm CP} \sim 500-650$~MeV indicating that heavy-ion collisions at energies $\sqrt{s_{\rm NN}} \simeq 3-7.7$~GeV could be most promising in the QCP CP search.
Existing experimental heavy-ion data do indicate excess of proton fluctuations around $\sqrt{s_{\rm NN}} \lesssim 10$~GeV as well as hints of non-monotonicity in $\kappa \sigma^2$. 
Future data from BES-II, STAR-FXT, and FAIR-CBM, coupled with the steadily improving dynamical modeling of (non-)critical fluctuations indicate that exciting times are ahead for the QCD CP search in the coming years.

\vskip3pt

{\bf Acknowledgments.}
The author thanks the organizers of Quark Matter 2023 for the invitation to give this plenary talk. The author also thanks A. Bzdak, M. Hippert, J. Karthein, V.~Koch, A. Pandav, P. Parotto, A. P\'asztor, R. Poberezhnyuk, M. Pradeep, K. Rajagopal, O.~Savchuk, and M.~Stephanov for fruitful discussions.

\bibliography{template}

\end{document}